\documentclass[twocolumn,english,aps,prb,floatfix,showpacs]{revtex4}
\usepackage{amsmath}
\usepackage{graphicx}
\usepackage{epstopdf}

\begin{document}

\title{Coarse graining tensor renormalization by the higher-order singular value decomposition}

\author{Z. Y. Xie$^1$, J. Chen$^2$, M. P. Qin$^2$, J. W. Zhu$^3$, L. P. Yang$^4$}
\author{T. Xiang$^{2,1}$}
\email{txiang@aphy.iphy.ac.cn}
\affiliation{$^{1}$Institute of Theoretical Physics, Chinese Academy of Sciences, P.O. Box 2735, Beijing 100190, China}

\affiliation{$^{2}$Institute of Physics, Chinese Academy of Sciences, P.O. Box 603, Beijing 100190, China}

\affiliation{$^3$Institute of Computational Mathematics and Scientific/Engineering Computing,
Academy of Mathematics and Systems Science, Chinese Academy of Sciences, Beijing 100190, China}

\affiliation{$^4$Beijing Computational Science Research Center, Beijing, 100084, China}

\date{\today}

\begin{abstract}
We propose a novel coarse graining tensor renormalization group method based on the higher-order singular value decomposition. This method provides an accurate but low computational cost technique for studying both classical and quantum lattice models in two- or three-dimensions. We have demonstrated this method using the Ising model on the square and cubic lattices. By keeping up to 16 bond basis states, we obtain by far the most accurate numerical renormalization group results for the 3D Ising model. We have also applied the method to study the ground state as well as finite temperature properties for the two-dimensional quantum transverse Ising model and obtain the results which are consistent with published data.
\end{abstract}

\pacs{05.10.Cc, 71.10.-w, 75.10.Hk}

\maketitle

\section{Introduction}

The simulation of two or higher dimensional quantum lattice models remains a great challenge. This has stimulated great interest on the investigation of renormalization group (RG) methods for the tensor-network states \cite{Nig-MPS, VersCirac-PEPS, Levin-TRG, Jianghongchen-BVP, Guzhengcheng-TERG, JOVVC-PEPSAlgor, Xiezhiyuan-SRG, Zhaohuihai-SRG, Chen2011, Vers-SinLayer, Nishino-CTTRG, Nishino-CTMRGKWA, Nishino-TPVA, Nishino-CTMRGVertex, Nishino-VDMA, Nishino-TPVAPara, Latoree-BondScheme, Wen-unpub, Roman-CTMReview}. The use of the tensor-network state as a variational wave function for the classical lattice model was first considered by Nishino and coworkers \cite{Nishino-CTMRGKWA, Nishino-TPVA, Nishino-CTMRGVertex, Nishino-VDMA, Nishino-TPVAPara}. They, and recently Garcia-Saez \textit{et.~al} \cite{Latoree-BondScheme}, proposed a number of RG approaches to study the thermodynamic properties of the Ising and other models. However, due to the heavy computational cost, the maximal truncated tensor dimension, $D$, that can be handled with their methods is small (between 2 and 5) in 3D, and consequently the accuracy of the results they obtained is low in comparison with the Monte Carlo ones.

In 2007, Levin and Nave \cite{Levin-TRG} proposed a coarse grained tensor renormalization group (TRG) method for studying two dimensional (2D) classical models based on the singular value decompostion (SVD) of matrix. Later we proposed a second renormalization group (SRG) method \cite{Xiezhiyuan-SRG, Zhaohuihai-SRG} to globally optimize the truncation scheme and improve significantly the accuracy of the TRG. The application of these methods in classical and quantum lattice models has achieved great success \cite{Jianghongchen-BVP, Guzhengcheng-TERG, Xiezhiyuan-SRG, Zhaohuihai-SRG, Chen2011}. However, it is difficult to extend these methods to 3D, not just due to the increase of the order of local tensors, but also due to the change of lattice topology in the coarse graining process \cite{Wen-unpub}.

In this paper, we introduce a novel coarse graining TRG method based on the higher-order singular value decomposition (HOSVD) \cite{Latheauwer-HOSVD} to study physical properties of 2D or 3D lattice models. We will first discuss about a simple TRG method based on the HOSVD (abbreviated as HOTRG hereafter), and then discuss about a more sophisticated method that incorporates the second renormalization effect of environment tensors to the HOTRG. The HOSVD takes into account more accurately the interplay between different components of a tensor. It provides a better scheme to truncate a local tensor than the SVD.

This paper is arranged as follows. In Sec.~\ref{sec:2D}, a detailed introduction to the HOTRG and HOSRG for the 2D statistical lattice models is given. We have taken the 2D Ising model to show how accurate the HOTRG and HOSRG can be in comparison with other methods. In Sec.~\ref{sec:3D}, we have extended the HOTRG and HOSRG to the 3D statistical lattice models. For the 3D Ising model, we have obtained by far the most accurate numerical renormalization group results for the 3D Ising model. Our accuracy is comparable with the best Monte Carlo results. In Sec.~\ref{sec:2D-Quantum}, we have applied HOTRG to 2D quantum lattice models. Our preliminary results show that the HOTRG provides a powerful tool for studying the ground state and thermodynamic properties of 2D quantum lattice models. A summary is given in Sec.~\ref{sec:Summary}.

\section{Two-dimensional systems}
\label{sec:2D}

\subsection{HOTRG}

\begin{figure}[tbp]
\includegraphics[width=0.8\columnwidth]{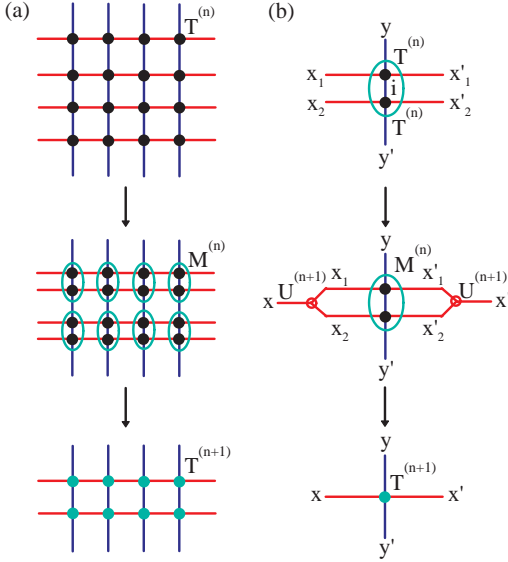}
\caption{
(a) A HOTRG contraction of the tensor network state along the y axis on the square lattice. (b) Steps of contraction and renormalization of two local tensors. The initial tensor $T^{(0)} = T$.
}
\label{fig:SquareRGflow}
\end{figure}

Let us start by taking the Ising model,
\begin{equation}
H = - \sum_{\langle ij \rangle }\sigma_z^i\sigma^j_z,
\end{equation}
as an example to show how the method works in 2D first. $\sigma_z^i$ is the pauli matrix at site $i$. An extension of the method to 3D will be described later. The partition function of the 2D Ising model can be represented as a translation invariant tensor network state \cite{Zhaohuihai-SRG},
\begin{equation}
Z = \mathrm{Tr}\prod_{i} T_{ x_{i}x'_{i}y_{i}y'_{i} } ,    \label{eq:PartFunc}
\end{equation}
where $i$ runs over all the lattice sites and $\mathrm{Tr}$ is to sum over all bond indices, and the local tensor $T$ is defined at each lattice site as shown in Fig.~\ref{fig:SquareRGflow}(a),
\begin{equation}
T_{x_{i}x'_{i}y_{i}y'_{i}} = \sum_{\alpha}W_{\alpha,x_i}W_{\alpha,x'_i}W_{\alpha,y_i}W_{\alpha,y'_i}
\end{equation}
where $W$ is a $2\times 2$ matrix defined by
\begin{equation}
W =
\left(
\begin{array}{cc}
\sqrt{\cosh (1/T) },  & \sqrt{\sinh (1/T) }  \\
\sqrt{\cosh (1/T) },  & -\sqrt{\sinh (1/T)}
\end{array}
\right),
\end{equation}
and $T$ is the temperature.

To coarse grain, we contract the lattice alternatively along the horizontal (x-axis) and vertical (y-axis) directions. This scheme of coarse graining is simple to implement. Fig.~\ref{fig:SquareRGflow}(a), as an example, shows how the contraction along the y-axis is done. At each step, two sites are contracted into a single site in the coarse grained lattice (Fig.~\ref{fig:SquareRGflow}(b)), and the lattice size is reduced by a factor of 2.

The contracted tensor at each coarse grained lattice site is defined by
\begin{equation}
M^{(n)}_{xx'yy'} = \sum_{i} T^{(n)}_{x_{1}x'_{1}yi} T^{(n)}_{x_{2}x'_{2}iy'} ,  \label{eq:2DFatTensor}
\end{equation}
where $x=x_1\otimes x_2$, $x'=x'_1 \otimes x'_2$, and the superscript $n$ denotes the $n$'th iteration. The bond dimension of $M^{(n)}$ along the x-axis is the square of the corresponding bond dimension of $T^{(n)}$. To truncate $M^{(n)}$ into a lower rank tensor, we first do a HOSVD for this tensor\cite{Latheauwer-HOSVD}
\begin{equation}
M^{(n)}_{xx'yy'} = \sum_{ijkl} S_{ijkl}U^L_{xi} U^R_{x'j} U^U_{yk} U^D_{y'l} , \label{eq:2DHOSVD}
\end{equation}
where $U$'s are the unitary matrices. $S$ is the core tensor of $M^{(n)}$, which possesses the following properties for any index, say index $j$:

(1) all orthogonality,
\begin{equation}
\langle S_{:,j,:,:} \mid S_{:,j',:,:} \rangle = 0, \qquad \mathrm{if} \,\, j\neq j',
\end{equation}
where $\langle S_{:,j,:,:} \mid S_{:,j',:,:} \rangle$ is the inner-product of these two sub-tensors.

(2) pseudo-diagonal,
\[
|S_{:,j,:,:}| \geq |S_{:,j',:,:}|, \qquad \mathrm{if} \,\, j < j',
\]
where $|S_{:,j,:,:}|$ is the norm of this sub-tensor which is the square root of all elements' square sum. These norms play a similar role as the singular values of a matrix.

In $M^{(n)}$, the two vertical bonds, $y$ and $y'$, do not need to be renormalized. Thus in the practical calculation, $U^U$ and $U^D$ are not needed to be determined. Moreover, the right bond of $M^{(n)}$ is linked directly to the left bond of an identical tensor on the right neighboring site, thus to truncate any one of the horizontal bonds of $M^{(n)}$ will automatically truncate the other horizontal bond. The truncation can be done by comparing the values of
\begin{equation}
\varepsilon_1 = \sum_{i>D} |S_{i,:,:,:}|^2
\end{equation}
and
\begin{equation}
\varepsilon_2 = \sum_{j>D} |S_{:,j,:,:}|^2.
\end{equation}
If $\varepsilon_1 < \varepsilon_2$, we truncate the first dimension of $S$ or the second dimension of $U^L$ to $D$. Otherwise, we truncate the second dimension of $S$ or the second dimension of $U^R$ to $D$. This kind of truncation scheme provides a simple and optimal approximation to minimize the truncation error\cite{Latheauwer-LowRank, Silva-LowRank}. It has been successfully applied to many fields such as data compression, image processing, pattern recognition, and etc \cite{HOSVD-AdvanApp}.

After the truncation, we can update the local tensor using the following formula
\begin{eqnarray}
T^{(n+1)}_{xx'yy'} = \sum_{ij} U^{(n+1)}_{ix} M^{(n)}_{ijyy'} U^{(n+1)}_{jx'} ,                                   \label{eq:2DTruncation}
\end{eqnarray}
where $U^{(n+1)} = U^L$ (or $U^R$) if $\varepsilon_1$ is smaller (or larger) than $\varepsilon_2$.

The above HOTRG calculation can be repeated iteratively until the free energy and other physical quantities calculated are converged. The cost of the calculation scales as $D^7$ in the computer time and $D^4$ in the memory space. This is comparable with the cost of TRG \cite{Xiezhiyuan-SRG, Zhaohuihai-SRG}.

The key step in the above HOTRG iteration is to determine the four unitary matrices on the right hand side of Eq.~(\ref{eq:2DHOSVD}). In our calculation, we determine these matrices by taking the singular value decomposition of matrices. As an example, let us consider how to evaluate $U^L$. We first convert $M_{xx'yy'}$ into a matrix $M^\prime_{x, \, x'yy'}$
\[
M^\prime_{x, \, x'yy'} = M_{xx'yy'}
\]
with the first index $x$ as the row index and the rest indices $(x',y,y')$ as the column index of this matrix. Then from the theory of HOSVD, we know that $U^{L}$ is equal to the left unitary matrix of $M$ under the singular value decomposition. Thus $U^{L}$ can be simply determined from the canonical transformation of the unitary matrix $M^\prime M^{\prime \dagger}$
\begin{equation}
M'M'^{\dagger} = U^{L}\Lambda^{L}(U^{L})^{\dagger},
\end{equation}
where $\Lambda^L$ is the eigenvalue of $M^\prime M^{\prime \dagger}$. Furthermore, it can be shown that
\begin{equation}
|S_{i,:,:,:}|^2 = \Lambda^{L}_i .
\end{equation}
The cost for evaluating these $U$-matrices scales with $D^{6}$.

\subsection{HOSRG}

The HOTRG is a local optimization method. It minimizes the error in the truncation of a local tensor. However, it ignores the renormalization effect of environment. To develop a global optimization method, it is necessary to consider the environment contribution in the renormalization of local tensors. In Refs.~\cite{Xiezhiyuan-SRG, Zhaohuihai-SRG}, we proposed a SRG appraoch to incorporate the environment contribution in the optimization of local tensors. This kind of SRG approach can be also used to improve the performance of HOTRG, which leads to a global optimized HOTRG method, referred as HOSRG below.

\begin{figure}[tbp]
\includegraphics[width=0.6\columnwidth]{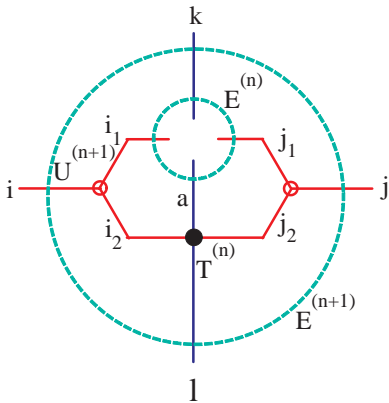}
\caption{(color online) Graphical representation of Eq.~(\ref{eq:2Diteration}) for determining the environment tensor $E^{(n)}_{kaj_{1}i_{1}}$ from $E^{(n+1)}_{ijkl}$ in the backward iteration.}
\label{fig:IterEnv}
\end{figure}

The HOSRG follows the same coarse graining steps as in the HOTRG. However, at each step, one needs to calculate a bond density matrix defined on a bond whose basis space will be truncated. This bond density matrix is defined by tracing out all environment tensors.

The SRG introduced in Ref.~\cite{Xiezhiyuan-SRG,Zhaohuihai-SRG} is an infinite lattice algorithm. In the calculation of the bond density matrix, the size of the environment is always assumed to be infinite. At each step of coarse graining, a combined forward and backward iteration is performed. In the forward iteration, the TRG is applied to determine all transformation matrices and local tensors. A backward iteration is then performed to determine the bond density matrix. This scheme can be readily extended to the HOSRG. We have done this kind of calculation and find that it does provide much more accurate results than the HOTRG.

Similar as in the DMRG, one can also introduce a finite lattice algorithm to perform the HOSRG calculation. In this case, the whole size of the system is fixed and the number of tensors in the environment is reduced at each step of coarse graining. Again the second renormalization effect of the environment is handled by performing forward-backward iterations. But now this kind of forward-backward iterations can be repeated for many times, similar as to do a finite size sweeping in the DMRG. This provides a self-consistent approach to treat the system as well as environment tensors. It can further improve the accuracy of the HOSRG. Below we give an introduction to this method.

In the first round of forward iteration, we carry out a standard HOTRG calculation to determine iteratively all the transformation matrices $U^{(n)}$ and local tensors $T^{(n)}$. This iteration ends when the system reaches a desired size, say $2^N$ lattice points with $N=30\sim 50$. One then carries out a backward iteration to calculate the environment tensor $E^{(n)}$ iteratively, starting from $E^{(N+1)}$ which is set to be an unit tensor. The iteration formula for determining $E^{(n)}$ is given by
\begin{equation}
E^{(n)}_{kaj_{1}i_{1}} = \sum_{ijli_{2}j_{2}a} E^{(n+1)}_{ijkl} T^{(n)}_{i_{2}j_{2}al} U^{(n+1)}_{i_{1}i_{2},i} U^{(n+1)}_{j_{1}j_{2},j}  . \label{eq:2Diteration}
\end{equation}
A graphical representation of this equation is shown in Fig.~\ref{fig:IterEnv}. This backward iteration is terminated after $E^{(2)}$ is determined.

\begin{figure}[tbp]
\includegraphics[width=0.65\columnwidth]{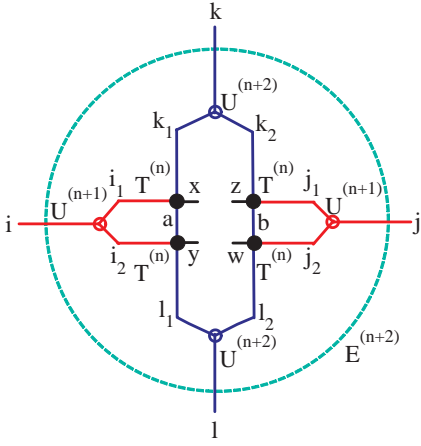}
\caption{(color online) Graphical representation of Eq.~(\ref{eq:bondDensityMatrix}) for determining the bond density matrix $\rho^{(n)}_{zw,xy}$ through $E^{(n+2)}_{ijkl}$.}
\label{fig:BondRDM}
\end{figure}

The above forward-backward iteration determines all coarse grained system and environment tensors that are needed for carrying out the second renormalization calculation. From these tensors we can perform another forward and backward iteration to improve their accuracy. But starting from this round of iteration, the HOTRG is no longer needed. Instead, at each step of coarse graining we evaluate the bond density matrix $\rho_{zw,xy}$ using the following formula
\begin{eqnarray}
\rho^{(n)}_{zw,xy} &=& E^{(n+2)}_{ijkl} U^{(n+1)}_{i_{1}i_{2}i} U^{(n+1)}_{j_{1}j_{2}j} U^{(n+2)}_{k_{1}k_{2}k} U^{(n+2)}_{l_{1}l_{2}l}
\nonumber \\
& & T^{(n)}_{i_{1}xk_{1}a} T^{(n)}_{i_{2}yal_{1}} T^{(n)}_{zj_{1}k_{2}b} T^{(n)}_{wj_{2}bl_{2}} .
\label{eq:bondDensityMatrix}
\end{eqnarray}
The repeated index summation is assumed. A graphical representation of this formula is shown in Fig.~\ref{fig:BondRDM}.

\begin{figure}[tbp]
\includegraphics[width=0.85\columnwidth]{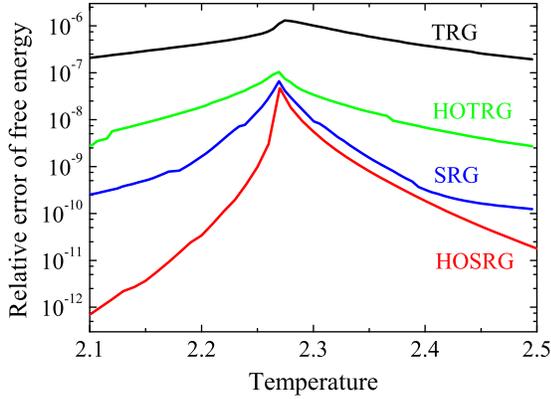}
\caption{(color online) Comparison of the relative errors of free energy with respect to the exact results for the 2D Ising model obtained by various methods with $D=24$. The critical temperature $T_c = 2 / \ln (1 + \sqrt{2}) $. }
\label{fig:SquareResult}
\end{figure}

To diagonalize this bond density matrix\cite{note}
\begin{equation}
\rho^{(n)} = U^{(n+1)} \Lambda \left( U^{(n+1)}\right)^{\dagger},
\end{equation}
we can find its eigenpair, $(\Lambda, U^{(n+1)})$. Same as in the density matrix renormalization group \cite{White-DMRG}, the eigenvalues of this density matrix determine the probabilities of the corresponding eigenvectors in the virtual bond basis space. By keeping the largest $D$ eigenvalues of $\Lambda$ and the corresponding eigenvectors of $U^{(n+1)}$, one can update the local tensor $T^{(n+1)}$ using Eq.~(\ref{eq:2DTruncation}). After finishing this forward iteration, we can take a backward iteration to update all environment tensors with Eq.~(\ref{eq:2Diteration}). This forward-backward iteration is then repeated until all system and environment tensors are converged.

Fig.~\ref{fig:SquareResult} compares the relative errors of free energy with respect to the rigorous solution \cite{OnSager-RigSqIsing} for the 2D Ising model obtained with four different methods. By keeping just 24 states, we find that the relative error of the HOTRG result is already less than $10^{-7}$ even at the critical temperature, much more accurate than the TRG result \cite{Xiezhiyuan-SRG, Zhaohuihai-SRG}. The HOSRG also performs better than the SRG. But the difference in the results obtained by these two methods is relatively small around the critical point. The HOTRG is less accurate than the two SRG methods, but it is computationally economic. The difference between TRG/SRG and HOTRG/HOSRG lies mainly in the basis truncation scheme. The former is based on the SVD, while the latter is based on the HOSVD. The above comparison indicates that the HOSVD scheme works better.

\begin{figure}[tbp]
\includegraphics[width=0.75\columnwidth]{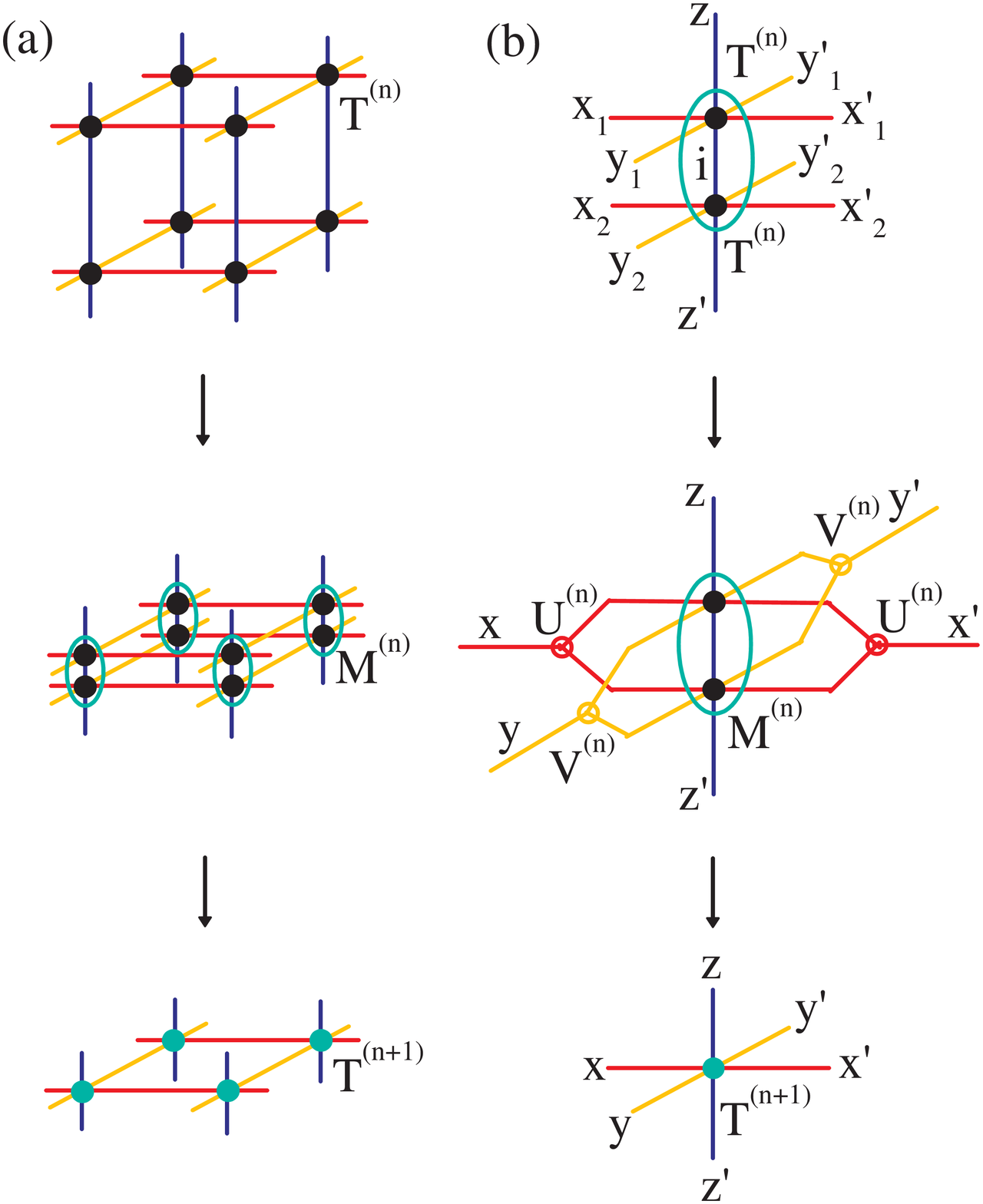}
\caption{(a) A HOTRG coarse graining step along the $z$-axis on the cubic lattice. (b) Steps of contraction and renormalization of two local tensors.}
\label{fig:Cubic-RGflow}
\end{figure}

\section{Three-dimensional systems}
\label{sec:3D}

The above HOTRG and HOSRG methods can be readily extended to 3D. This is an advantage of the coarse graining scheme proposed here. On the cubic lattice, a full cycle of lattice contraction needs to be done in three steps, along the x-axis, y-axis, and z-axis, respectively. At each step, two neighboring tensors will be combined to form a single coarse grained tensor and the lattice size is reduced by a factor of 2.

As an example, Fig.~\ref{fig:Cubic-RGflow} shows how the tensors are contracted along the z-axis. The HOSVD of the coarse grained local tensor (Fig.~\ref{fig:Cubic-RGflow}(b)) can be similarly done as for the 2D case. But the local tensor now has six bond indices and a HOSVD for a higher order tensor should be done. Moreover, the basis spaces for both the x-axis and y-axis bonds need to be renormalized. Thus we should determine from the core tensor and the unitary matrices of $M^{(n)}$ not only the transformation matrix for the x-direction bonds, $U^{(n)}$, but also the transformation matrix for the y-direction bonds, $V^{(n)}$. After that the dimensions for both x-axis and y-axis bonds are truncated and the local tensor is updated using $U^{(n)}$ and $V^{(n)}$. The contraction and renormalization of tensors along other two directions can be similarly done. This three-step iteration can then be repeated until the results are converged.

\begin{figure}[tbp]
\includegraphics[width=0.6\columnwidth]{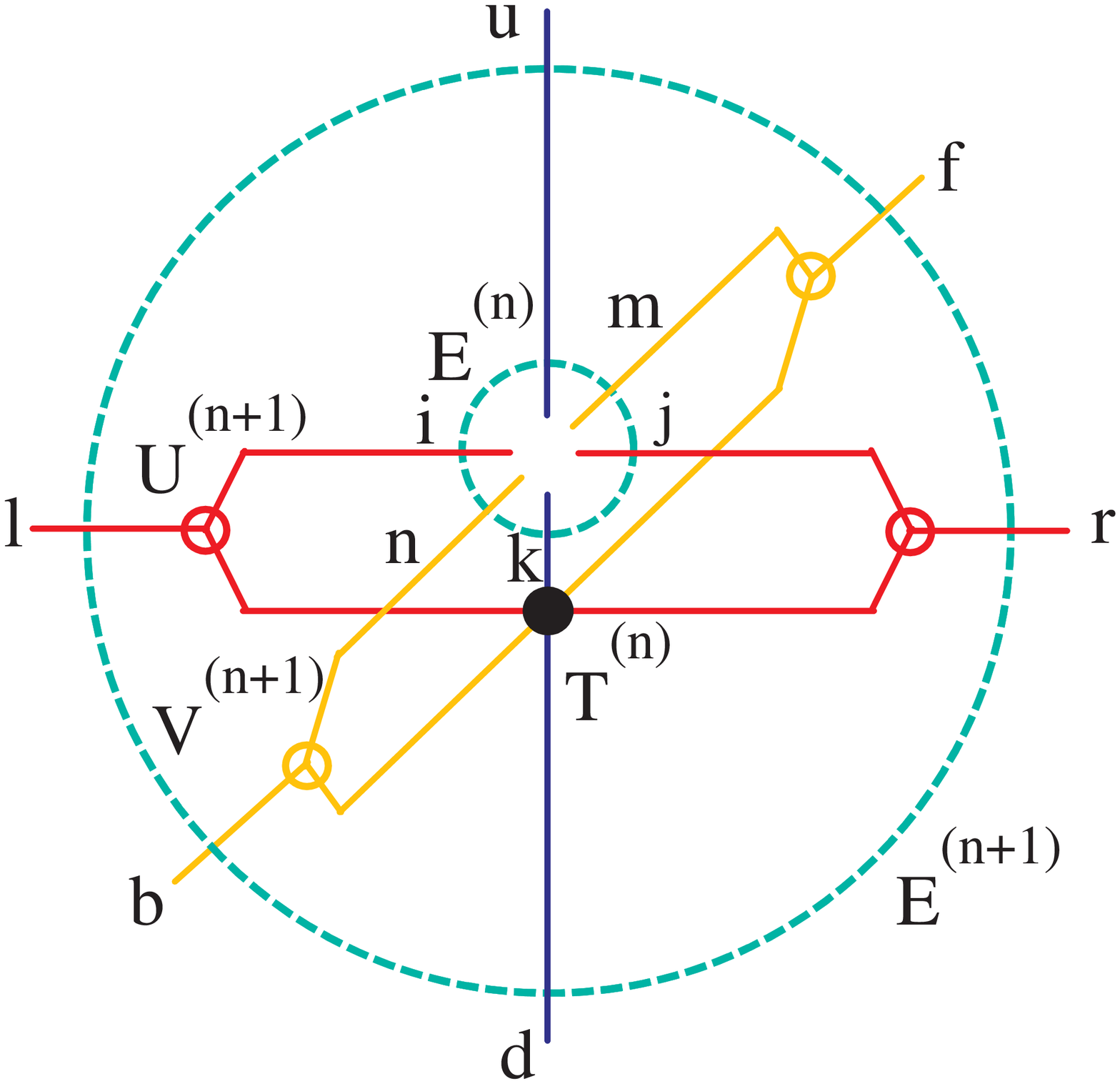}
\caption{(Color online) Graphical representation for the determination of the environment tensor $E^{(n)}_{mnjiuk}$ from $E^{(n+1)}_{lrfbud}$ in 3D.}
\label{fig:3DIterEnv}
\end{figure}

\begin{figure}[bp]
\includegraphics[width=0.75\columnwidth]{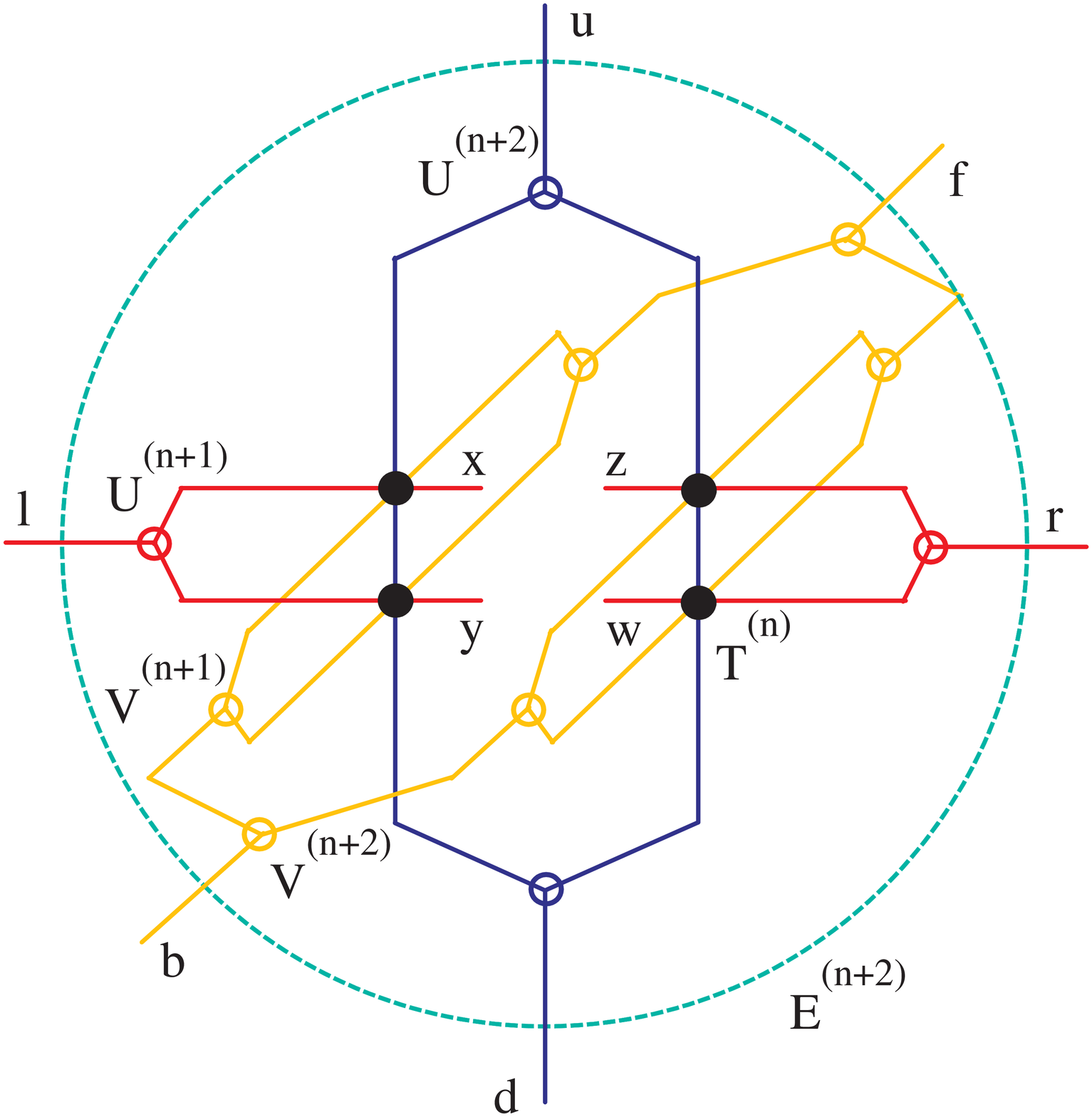}
\caption{(Color online) Graphical representation for the determination of the bond density matrix $\rho^{(n)}_{zw,xy}$ from the environment tensor $E^{(n+2)}_{lrfbud}$ in 3D.}
\label{fig:3DBondRDM}
\end{figure}

After the above HOTRG iteration, one can also do a backward iteration to evaluate the environment tensors and carry out the HOSRG calculation in 3D. A graphical representation for iteratively determining the environment tensor in this backward iteration is shown in Fig.~\ref{fig:3DIterEnv}. A series of forward-backward iterations is then performed to take into account the second renormalization effect of the environment to the coarse grained tensors. In the subsequent forward iterations, we evaluate and diagonalize the bond density matrix (see Fig.~\ref{fig:3DBondRDM}) and update the coarse grained tensors. The environment tensors are evaluated again in the backward iteration.

In the 3D calculation, the computational time scales with $D^{11}$ and the memory scales with $D^6$. This cost in the computational resource is significantly smaller than in other 3D numerical RG methods \cite{Nishino-CTTRG, Nishino-CTMRGKWA, Nishino-TPVA, Nishino-CTMRGVertex, Nishino-VDMA, Nishino-TPVAPara, Latoree-BondScheme, Roman-CTMReview}. We have studied the 3D Ising model using the HOTRG for $D$ up to 16.

\begin{figure}[tbp]
\center{\includegraphics[angle=0,width=\linewidth]{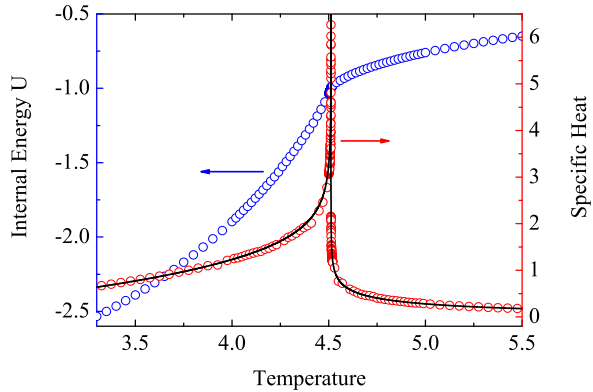}}
\caption{(Color online) The internal energy and the specific heat for the 3D Ising model obtained by the HOTRG with D = 14. The Monte Carlo result (black curve) obtained from an empirical fit formula given in Ref.~[\onlinecite{XiaomeiFeng-HeatCurveEc}] is shown for comparison.}
\label{HOTRGEnergy}
\end{figure}

\begin{table}[b]
\caption{Comparison of the internal energy at the critical temperature $U_c$ for the 3D Ising model obtained by different methods. }
\vspace{-12pt}
\begin{center}
\def\temptablewidth{0.45\textwidth}{\rule{\temptablewidth}{1pt}}
\begin{tabular*}{\temptablewidth}{@{\extracolsep{\fill}}llcc}\hline
method & $\qquad U_{c}$  \\
\hline
HOTRG(D = 16) & -0.990842(3) \\
Series expansion \cite{Comi-SeriExpEc} &  -0.991(1)  \\
Series expansion \cite{Fisher-SeriExpEc} &  -0.9902(1) \\
Series expansion \cite{Hunter-SeriExpEc} &  -0.99218(15)  \\
Monte Carlo \cite{XiaomeiFeng-HeatCurveEc} & -0.990604(4) \\
Monte Carlo \cite{Pinn-MCEc} & -0.9904(8) \\
Monte Carlo \cite{Jensen-MCEc} & -0.990(4) \\
\hline
\end{tabular*} \label{Tab:Uc}
{\rule{\temptablewidth}{1pt}}
\end{center}
\end{table}

\begin{figure}[tbp]
\center{\includegraphics[angle=0,width=\linewidth]{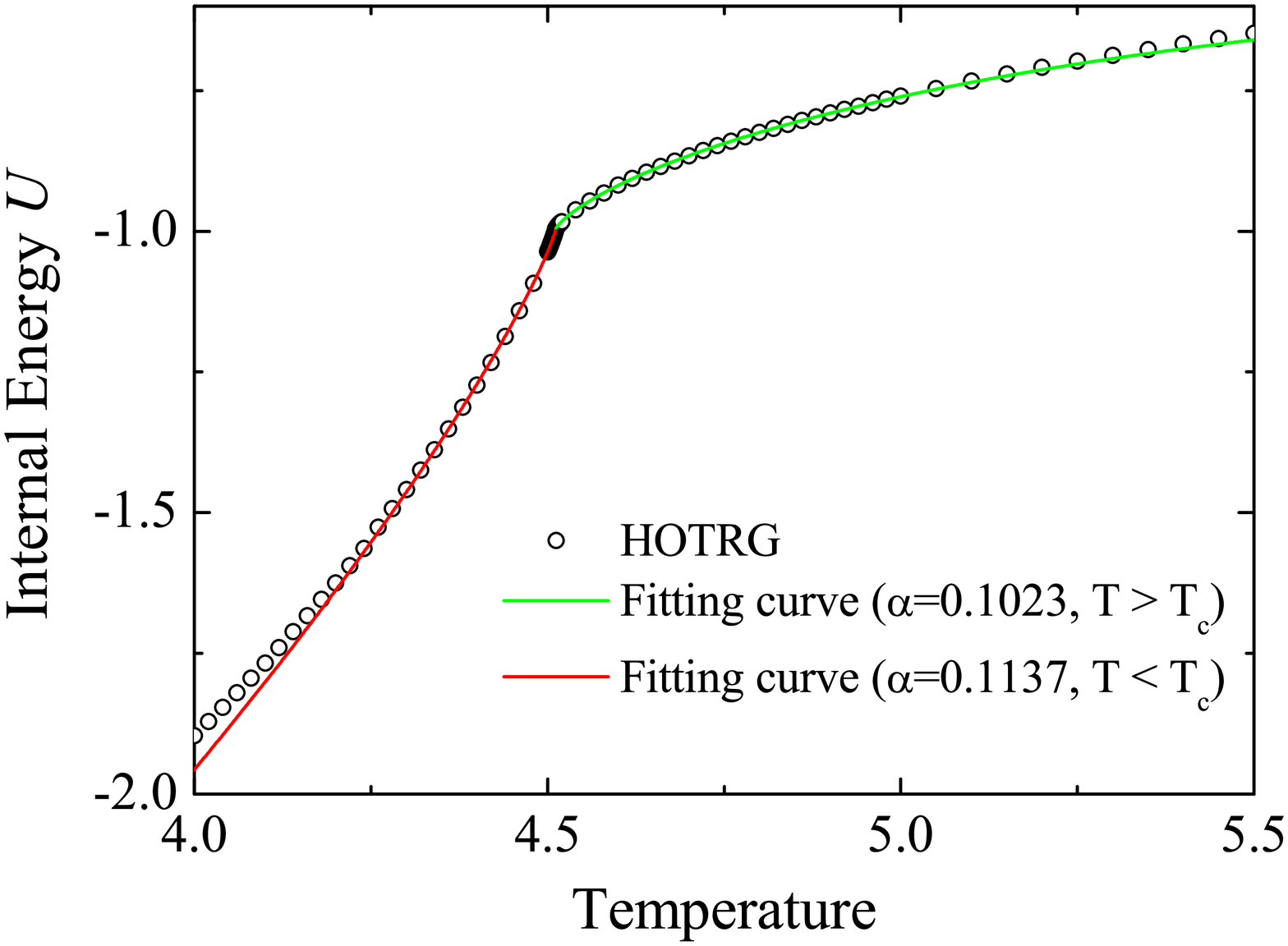}}
\caption{(Color online) The internal energy ($D=14$) and its fitting curves with Eq.~(\ref{eq:Ufit}) around the critical point for the 3D Ising model. $\alpha$ is the critical exponent for the specific heat.}
\label{HOTRGEnergyfit}
\end{figure}

\begin{figure}[bp]
\center{\includegraphics[angle=0,width=\linewidth]{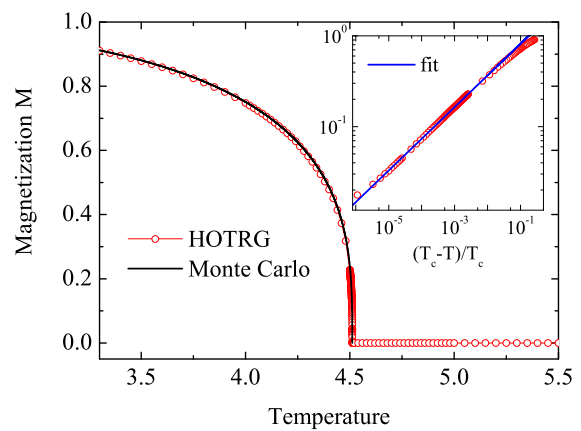}}
\caption{(color online) Temperature dependence of the mag-
netization for the 3D Ising model (D = 14). The Monte Carlo
result is from Ref. [\onlinecite{Talapov-SponFormulaTc}]. Inset: Logarithmic plot the magnetization around the critical point. The slope of the fitting curve
gives the critical exponent of the magnetization, $\gamma=0.3295$.}
\label{HOTRGSponM}
\end{figure}

The temperature dependence of the internal energy $U$ and the specific heat $C$ for the 3D Ising model obtained by the HOTRG with $D = 14$ is shown in Fig.~\ref{HOTRGEnergy}  and compared with the Monte Carlo result\cite{XiaomeiFeng-HeatCurveEc}. Our result of the specific heat agrees with the Monte Carlo one. At the critical temperature, $T_{c}=4.511544$, the internal energy is found to be $U_{c}=-0.995592$ for $D=14$. This value of $U_c$, as shown in Table \ref{Tab:Uc}, also agrees well with other published data.

>From the temperature dependence of the specific heat around the critical point, one can estimate the critical exponent of the specific heat with the formula
\begin{equation}
C \sim t^{-\alpha}  \label{eq:AlphaExp}
\end{equation}
where $t = |1 -T/T_{c}|$. However, as the specific heat data are obtained simply from the numerical derivative of the internal energy, the accuracy of the specific heat data is much less than that of the internal energy, especially around the critical point. This causes a big error in the determination of the exponent $\alpha$ with the above formula. This problem can be solved by directly evaluating this exponent from the temperature dependence of the internal energy. From the temperature integration of the specific heat, it is simple to show that the internal energy should exhibit the following critical behavior
\begin{equation}
U =  U_{c} + a t + b t^{1-\alpha} ,
\label{eq:Ufit}
\end{equation}
where $a$ and $b$ are unknown parameters which can be determined by fitting.

Fig.~\ref{HOTRGEnergyfit} shows the fitting curves for the internal energy around the critical point obtained with Eq.~(\ref{eq:Ufit}). The critical exponent is found to be $\alpha =0.1023$ and $0.1137$ for the temperature higher and lower than the critical value, respectively. These values of the critical exponent are consistent with the result obtained from the series expansion\cite{Fuji-SeriExpAlpha}, 0.104, and the Monte Carlo calculation\cite{Martin-MCAlphaBeta}, 0.111.

\begin{figure}[tbp]
\center{\includegraphics[angle=0,width=\linewidth]{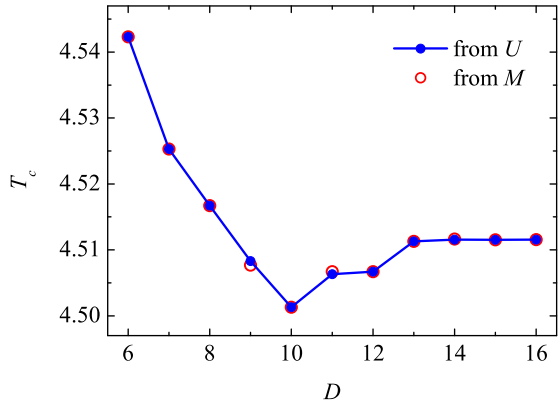}}
\caption{(Color online) The critical temperature $T_c$ as a function of the bond dimension $D$ for the 3D Ising model obtained from the internal energy ($U$) and magnetization ($M$), respectively.}
\label{HOTRGDimAnalyze}
\end{figure}

\begin{table}[bp]
\caption{Comparison of the critical point $T_{c}$ for the 3D Ising model obtained by different methods. }
\vspace*{-12pt}
\begin{center}
\def\temptablewidth{0.45\textwidth}
{\rule{\temptablewidth}{1pt}}
\begin{tabular*}{\temptablewidth}{@{\extracolsep{\fill}}llcc}\hline
$\quad$ method & $\quad T_{c}$  \\
\hline
HOTRG(D=16, from U)  &  4.511544   \\
HOTRG(D=16, from M)  &  4.511546   \\
Monte Carlo \cite{Martin-MCTc} &  4.511523  \\
Monte Carlo \cite{Dengyoujin-MCTc} &  4.511525  \\
Monte Carlo \cite{Gupta-MCTc} & 4.511516 \\
Monte Carlo \cite{Talapov-SponFormulaTc} &  4.511528 \\
Series expansion \cite{Comi-SeriExpTc} & 4.511536 \\
CTMRG \cite{Nishino-CTMRGKWA} & 4.5788 \\
TPVA \cite{Nishino-TPVA} & 4.5704 \\
CTMRG \cite{Nishino-CTMRGVertex}  & 4.5393 \\
TPVA \cite{Nishino-TPVAPara}  & 4.554 \\
Algebraic variation \cite{Chung-Tc} &  4.547  \\
\hline
\end{tabular*} \label{Tab:Tc}
{\rule{\temptablewidth}{1pt}}
\end{center}
\end{table}

Fig.~\ref{HOTRGSponM} shows the temperature dependence of the spontaneous magnetization $M$ obtained by the  HOTRG with D=14. Our data agree well with the Monte Carlo results\cite{Talapov-SponFormulaTc}. From the singular behavior of $M$, we find that the critical temperature $T_{c} = 4.511615$ for $D=14$. Furthermore, by fitting the data of $M$ in the critical regime with the formula
\begin{equation}
M \sim t^{\gamma} ,
\end{equation}
we find that the exponent $\gamma =0.3295$, consistent with the Monte
Carlo\cite{Martin-MCAlphaBeta} (0.3262) and series expansion\cite{Vicari-SeriExpBeta} (0.3265) results.

Fig.~\ref{HOTRGDimAnalyze} shows the critical temperature $T_c$ determined from the singular points of the internal energy as well as the magnetization for $D$ up to 16. The values of $T_c$ obtained from these two quantities agree with each other.
For $D = 16$, Tc obtained from the internal energy and the magnetization are 4.511544 and 4.511546, respectively. The relative difference is less than $10^{-6}$. But $T_c$ does not vary monotonically with $D$. It becomes converged
only when $D \ge 13$, indicating the importance of keeping a large $D$ in the 3D TRG calculation. The error in $T_c$, estimated from the difference between the values of $T_c$ for $D = 15$ and $D = 16$, is also less than $10^{-6}$. Our results agree with the Monte Carlo data\cite{Martin-MCTc,Dengyoujin-MCTc,Gupta-MCTc}.

The above discussion indicates that the HOTRG works very well in 3D. The accuracy of the results can be further improved by applying the HOSRG. However, the HOSRG calculation costs much more CPU time. A thorough study with the HOSRG on the 3D Ising model is still in progress and the results will be published separately.

\section{Ground State and Thermodynamics of 2D Quantum Lattice Models}
\label{sec:2D-Quantum}

A $d$-dimensional quantum lattice model is equivalent to a $(d+1)$-dimensional classical model, the HOTRG and HOSRG methods above introduced can be also extended to study the ground state and thermodynamic properties of $d$-dimensional quantum lattice models. For one-dimensional quantum lattice models, there are already many mature methods for studying the ground state as well as the thermodynamic properties. For example, the ground state can be studied by the DMRG\cite{White-DMRG} and the thermodynamics can be studied by the quantum transfer matrix renormalization group (TMRG)\cite{Bursill96,Wang97}. Here we will only discuss how to apply the HOTRG/HOSRG to a 2D quantum lattice model.

\begin{figure}[tbp]
\center{\includegraphics[angle=0,width=\linewidth]{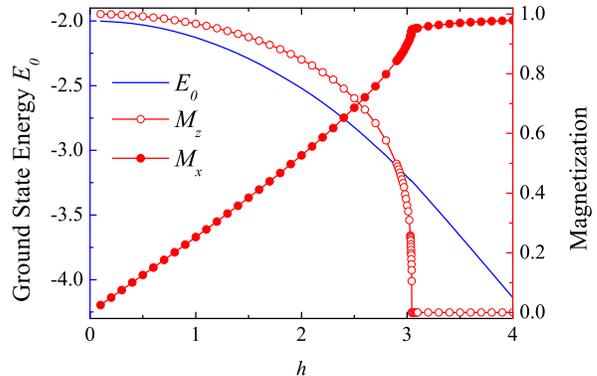}}
\caption{(Color online) The ground state energy $E_0$, the magnetization $M_{x}\equiv\langle \sigma_{x}\rangle$ and $M_{z}\equiv\langle \sigma_{z}\rangle$ versus the applied field $h$ for the 2D quantum Ising model obtained by the HOTRG  with $D=14$.}
\label{QisingGrd}
\end{figure}

\begin{table}[b]
\caption{The critical field $h_{c}$ for the 2D quantum Ising model with transverse field obtained by different methods. }
\vspace*{-12pt}
\begin{center}
\def\temptablewidth{0.45\textwidth}
{\rule{\temptablewidth}{1pt}}
\begin{tabular*}{\temptablewidth}{@{\extracolsep{\fill}}llcc}
\hline
$\quad$ method & $\,\, h_{c}$  \\
\hline
 HOTRG(D=14)  &  3.0439   \\
 Monte Carlo \cite{Dengyoujin-MChc} &  3.044  \\
 Series Expansion  \cite{Hamer-SeriExphc} &  3.044 \\
 iPEPS \cite{JOVVC-PEPSAlgor} &  3.06  \\
 VDMA \cite{Nishino-VDMA} & 3.2 \\
 TERG \cite{Guzhengcheng-TERG} & 3.08 \\
 iPEPS \cite{Roman-iPEPShc} & 3.04 \\
 CTM \cite{Roman-CTMReview}  & 3.14 \\
\hline
\end{tabular*} \label{hc}
{\rule{\temptablewidth}{1pt}}
\end{center}
\end{table}

As an example, we will take the 2D quantum Ising model with a transverse field to show how these methods work. The Hamiltonian of this model is defined by
\begin{equation}
H = -\sum_{\langle ij\rangle}\sigma^{i}_{z}\sigma^{j}_{z} - h\sum_{i}\sigma^{i}_{x}.
\end{equation}
We start by representing the partition function of this model as a tensor-network model in the 2+1 dimensions. By using the Trotter-Suzuki decomposition formula, we can express the partition function as\cite{Roman-CTMReview}
\begin{equation}
Z  =  \mathrm{Tr} e^{- \beta H} \approx  \mathrm{Tr} \left[  e^{- \tau H_z} e^{- \tau H_x} \right]^L + O(\tau^2)
\label{eq:Trotter}
\end{equation}
where
\begin{eqnarray}
H_z & = & -\sum_{\langle ij\rangle}\sigma^{i}_{z}\sigma^{j}_{z} ,\\
H_x & = & - h\sum_{i}\sigma^{i}_{x}.
\end{eqnarray}
$\beta = L \tau$ is the inverse temperature and $\tau$ is a small Trotter parameter. This partition function can be also expressed as a product of evolution matrix $V$
\begin{equation}
Z = \mathrm{Tr} V^L,
\end{equation}
where
\begin{equation}
V = e^{-\tau H_x/2} e^{-\tau H_z} e^{-\tau H_x/2}
\end{equation}
is the evolution operator between two neighboring Trotter layers. To insert the complete basis set between any two of the exponential terms on the right hand side of Eq.~(\ref{eq:Trotter}), it is straightforward to show that $V$ can be expressed as a product of local tensors.  From this, we can express the partition function as a 3D tensor-network model
\begin{equation}
Z \approx \mathrm{Tr} \prod_i T_{l_ir_if_ib_i\mu_i\nu_i} + O(\tau^2) ,
\label{eq:partition}
\end{equation}
where the six-indexed local tensor is defined by \begin{equation}
T_{lrfb\mu\nu} = \sum_{\sigma}W'_{\sigma l} W'_{\sigma r} W'_{\sigma f} W'_{\sigma b} P_{\sigma\mu} P_{\sigma\nu} ,
\end{equation}
with
\begin{equation}
W' = \left(
\begin{array}{cc}
\sqrt{\cosh{\tau}} ,  & \sqrt{\sinh{\tau}}  \\
\sqrt{\cosh{\tau}} , & -\sqrt{\sinh{\tau}}
\end{array}
\right) ,
\end{equation}
and
\begin{equation}
P = \frac{1}{\sqrt{2}}
\left(
\begin{array}{cc}
e^{\tau h/2} , & e^{-\tau h/2}  \\
e^{\tau h/2}  , & -e^{-\tau h/2}
\end{array}
\right) .
\end{equation}

The operator $V$ governs the basis state evolution along the Trotter direction. The matrix element of $V$ between two sets of basis states, $\langle \{ \nu_i \} |$ and $| \{ \mu_i \} \rangle$, is defined by tracing out all spatial indices of local tensors $T$ in a given Trotter layer
\begin{equation}
\langle \{ \nu_i \} | V | \{ \mu_i \} \rangle \equiv \mathrm{Tr}^\prime \prod_i
T_{l_ir_if_ib_i\mu_i\nu_i},
\end{equation}
where $\mathrm{Tr}^\prime $ is to trace out all spatial indices only.

\begin{figure}[tbp]
\center{\includegraphics[angle=0,width=\linewidth]{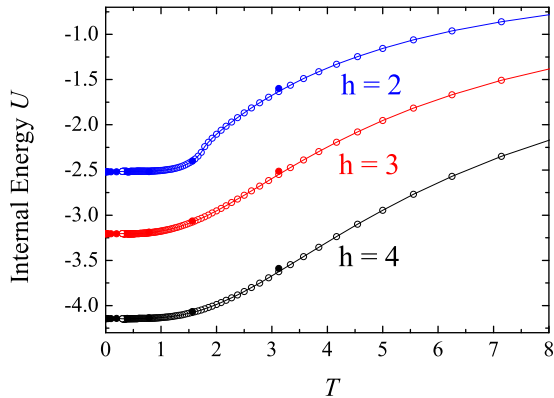}}
\caption{(Color online) Temperature dependence of the internal energy in three different fields. The solid dots and open circles are obtained with the 3D coarse graining HOTRG with $D=14$ and the imaginary time evolution approach with $D=24$, respectively. }
\label{QisingEnergy}
\end{figure}

\begin{figure}[tbp]
\center{\includegraphics[angle=0,width=\linewidth]{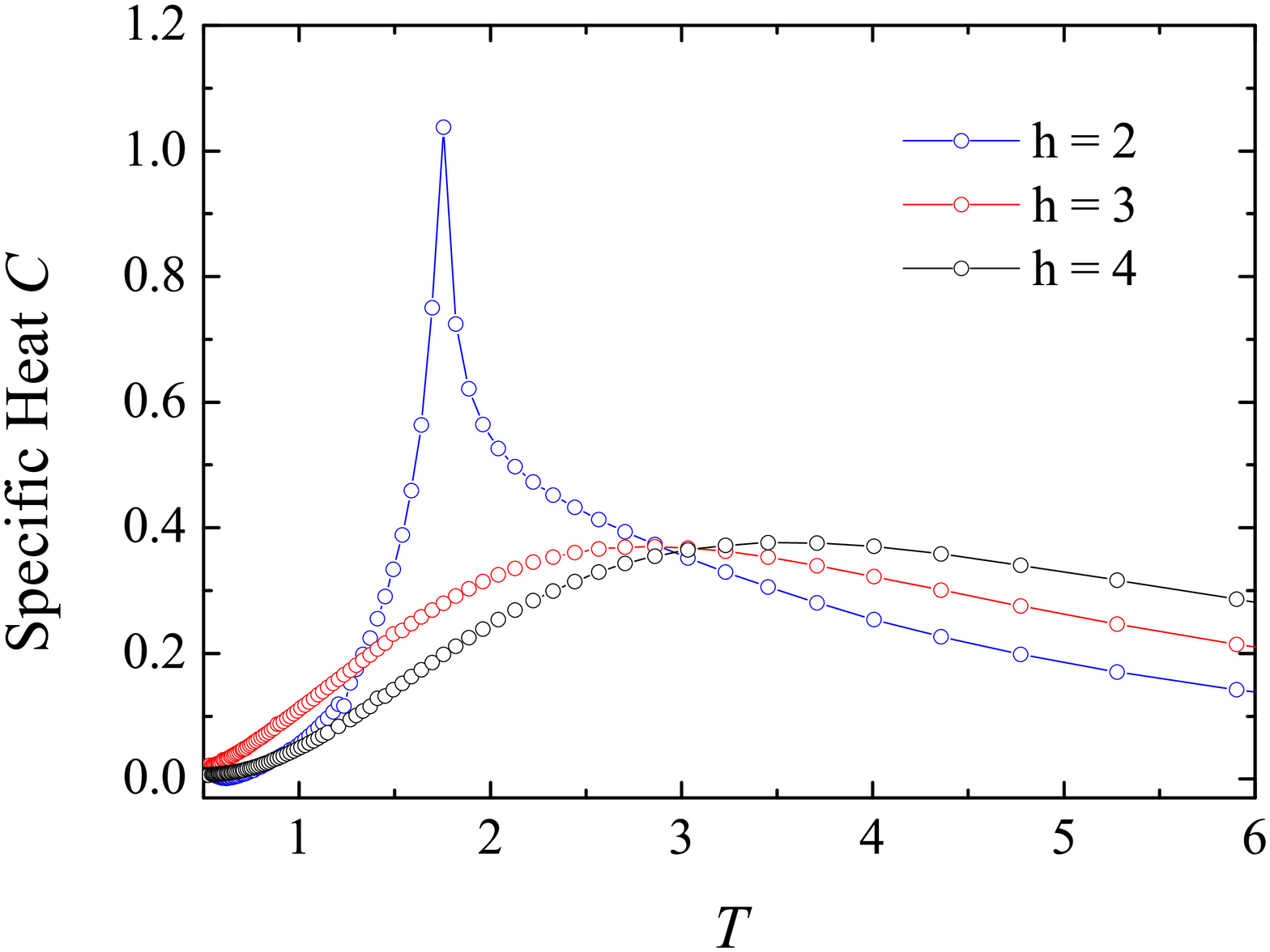}}
\caption{(Color online) The specific heat versus temperature obtained by the imaginary time evolution approach with $D=24$ for the 2D quantum Ising model. }
\label{QisingHeat}
\end{figure}

The transverse field $h$ appears only in the $P$-matrix. When $h=0$, it is simple to show that $PP^\dagger = I$. In this case, the Hamiltonian returns to the 2D classical Ising model and Eq.~(\ref{eq:Trotter}) is reduced to a product of $L$ 2D tensor-network model for the Ising model with an inverse temperature $\tau$.

At zero temperature, $L\rightarrow \infty$, the partition function is a product of infinite many tensors along all three directions. The 3D HOTRG or HOSRG method introduced previously can be directly applied to study physical properties of the ground state. In this case, there is no need to evaluate the ground state wavefunction. This is an advantage of this approach in comparison with other methods which are based on the tensor-network representation of ground state wavefunction.

Fig.~\ref{QisingGrd} shows the field dependence of the ground state energy $E_{0}$, the transverse magnetization $M_{x}$ and the longitudinal magnetization $M_{z}$ for the transverse Ising model obtained by the HOTRG with $D=14$ and $\tau = 0.01$. In agreement with other calculations, we find that this model exhibits a phase transition at a finite field. The critical field is found to be $h_{c}=3.0439$, consistent with other published data, as shown in Table~\ref{hc}.

At finite temperature, the lattice dimension of the tensor network model is finite along the imaginary time direction. Now two approaches can be used to evaluate the partition function. The first is to follow the steps of 3D HOTRG to contract the tensors alternatively along the three directions. The contraction along the Trotter direction is terminated if all bond variables along that direction are contracted. The iteration is then carried out purely along the two spatial directions as for a pure 2D classical model. This is an accurate and efficient approach for evaluating physical quantities. However, the number of temperature points that can be studied with this approach is quite limited for a given $\tau$, since the temperature is reduced by a factor of 2 at each contraction along the Trotter direction.

The second approach is to do the imaginary time evolution. In this case, one starts from a one-Trotter layer tensor product system, whose tensor operator is defined by $V$. At each step of evolution, one more Trotter layer (i.e. one $V$ operator) is added to the system and the truncation for the spatial bond dimension is done using the HOTRG.  By performing this imaginary time evolution up to certain temperature, physical quantities are then evaluated at that temperature by tracing out all bond variables with the HOTRG. The inverse temperature increases linearly with the evolution number. This allows us to collect more temperature points. However, as the truncation error is accumulated with the evolution, the results such obtained would become less and less accurate with decreasing temperature. Thus this approach should be used only for evaluating thermodynamic quantities in high temperatures. A similar imaginary time evolution approach was recently proposed for evaluating thermodynamic quantities by Ran et al.\cite{Ran2012} based on the bond entanglement mean field approach proposed in Ref. \cite{Jianghongchen-BVP}.

In our calculation, both approaches have been used. The first approach (i.e. the 3D coarse graining HOTRG) is more accurate than the second one (i.e. the imaginary time evolution approach) especially in low temperatures. It is used for collecting the low temperature data. The second approach allows more temperature points to be evaluated and is applied to evaluate thermodynamic quantities in high temperatures.

\begin{figure}[tbp]
\center{\includegraphics[angle=0,width=\linewidth]{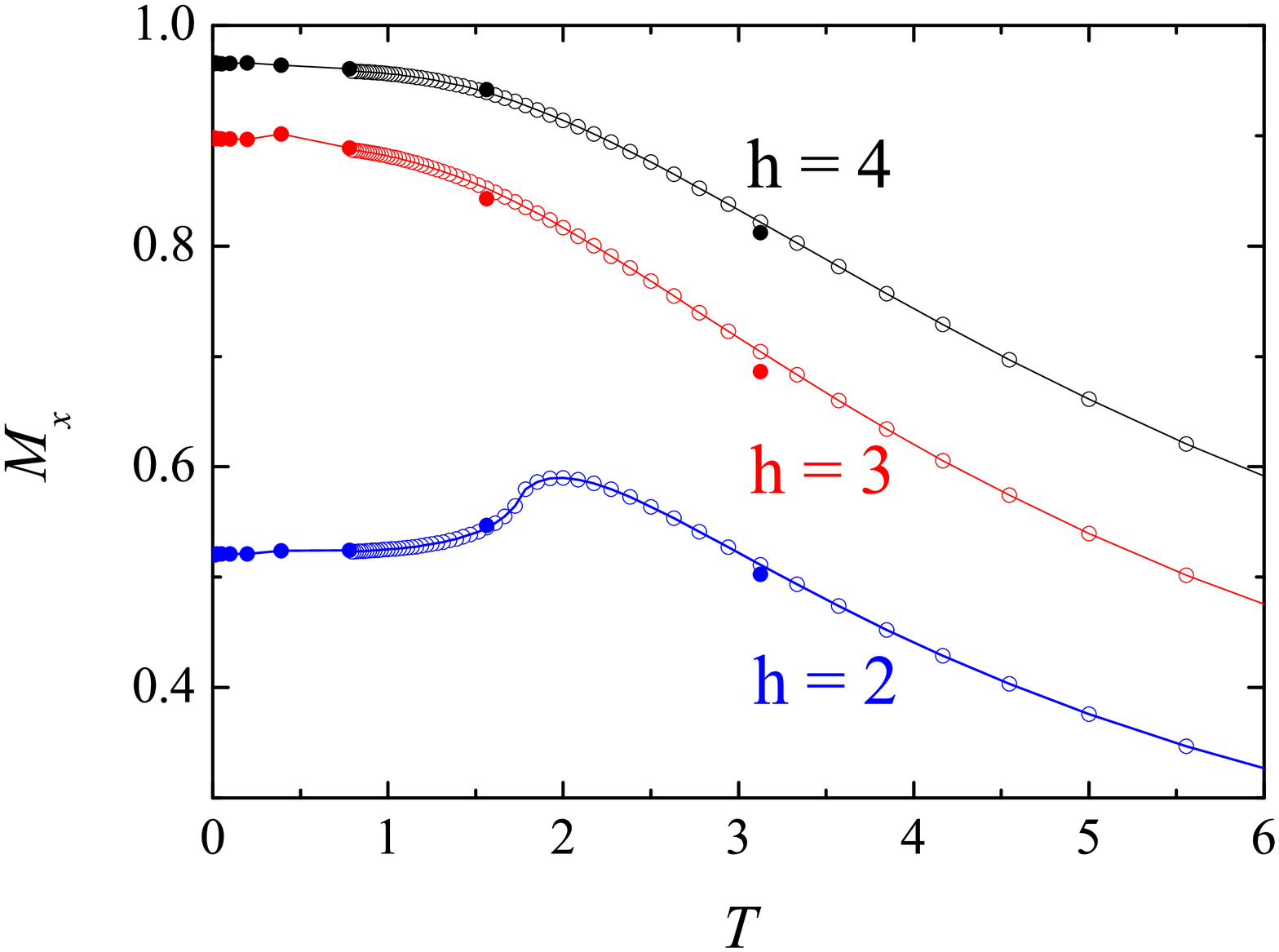}}
\caption{(Color online) Transverse magnetization as a function of temperature for the 2D quantum Ising model. The solid dots and open circles are obtained with the 3D coarse graining HOTRG with $D=14$ and the imaginary time evolution approach with $D=24$, respectively. }
\label{QisingMx}
\end{figure}

\begin{figure}[tbp]
\center{\includegraphics[angle=0,width=\linewidth]{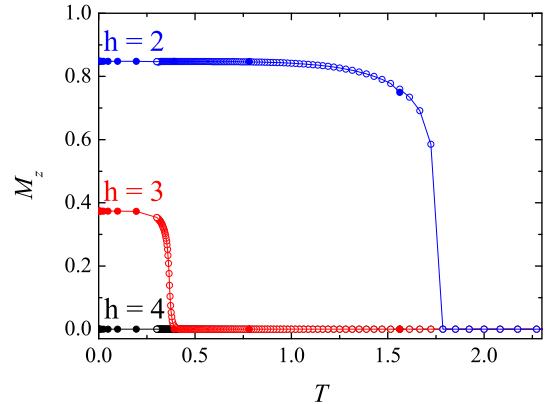}}
\caption{(Color online) Longitudinal magnetization as a function of temperature for the 2D quantum Ising model. The solid dots and open circles are obtained with the 3D coarse graining HOTRG with $D=14$ and the imaginary time evolution approach with $D=24$, respectively. }
\label{QisingMz}
\end{figure}

Figs.~\ref{QisingEnergy}-\ref{QisingMz} show the internal energy, the specific heat, the transverse and longitudinal magnetization as a function of temperature for the 2D quantum Ising model in three different applied fields, respectively. The solid and open circles are results obtained by the 3D HOTRG ($D=14$) and the imaginary time evolution ($D=24$) approaches, respectively. The results obtained with these two approaches agree with each other in the intermediated temperature range. In higher temperature, the truncation error of the imaginary time evolution is relatively small and the results obtained with this method is more accurate. In low temperature the results obtained by the 3D HOTRG are much more accurate than those obtained by the imaginary time evolution, since the error is accumulated at every step of coarse graining or time evolution and the 3D HOTRG can reach low temperatures in a much fewer coarse graining steps. In all calculations, the Trotter step is set to $\tau = 0.01$.

When $h < h_c$, as the ground state is spontaneously symmetry broken with a finite magnetic order, a finite temperature phase transition is expected. Such phase transition is confirmed by our calculation, which can be clearly seen from the temperature dependence of the longitudinal magnetization $M_z$ (Fig.~\ref{QisingMz}). The critical temperature decreases with increasing field $h$ for $h < h_c$. The specific heat is obtained from the numerical derivative of the internal energy shown in Fig.~\ref{QisingEnergy}. It shows a singular behavior around the critical point for $h=2$.

The above discussion indicates that the HOTRG (including the imaginary time evolution approach) provides a simple and powerful method for studying the ground state and finite temperature properties of 2D quantum lattice models. By applying the HOSRG to the above calculation, we believe that the accuracy of results can be further improved. But this takes a longer time to do the calculation.

\section{Summary}
\label{sec:Summary}

In summary, we have proposed the HOTRG and HOSRG methods for studying physical properties of classical or quantum lattice models in 2D or 3D. By comparison with the exact solution of the 2D Ising model, we have shown that the simple HOTRG calculation can already produce very accurate numerical results. The HOSRG takes into account the second renormalization effect of the environment tensors, and it can significantly improve the accuracy of the HOTRG. These method allow us to retain an unprecedentedly high bond dimension in the basis truncation and yield by far the most accurate numerical RG results for the 3D Ising model. We have also applied the HOTRG to study the ground state and thermodynamic properties of the 2D quantum Ising model with a transverse magnetic field. Our results agree well with all published data. Symmetry or good quantum number of the tensor network state for these models can be used to reduce the computational and storage cost. This will allow us to retain more basis states to reduce the truncation error and further improve the accuracy of results.

In this work, we have taken two translation invariant tensor network models, namely the Ising model and the transverse Ising model, as examples to show how the methods work. However, it should be emphasized that our methods work more generally. They can be extended to a system which is translation invariant by shifting two or more lattice sites, or even to a random system, such as a spin glass model. By combining with the local update method of quantum tensor product wavefunction introduced in Ref.~\onlinecite{Jianghongchen-BVP}, one can also use this method to study ground state properties of 3D quantum lattice models.

We wish to thank H.W.J. Blote for sending us the Monte Carlo result shown in Fig.~\ref{HOTRGEnergy}. We are indebted to Prof. T. Nishino for helpful discussion. This work was supported by NSFC (Nos. 10934008 and 10874215) and MOST 973 Project (No. 2011CB309703).


\begin{thebibliography}{99}


\bibitem{Nig-MPS} H. Niggemann, A. Klumper, and J. Zittartz, Z. Phys. B \textbf{104}, 103 (1997).

\bibitem{VersCirac-PEPS} F. Verstraete and J. Cirac, arXiv:0407066.

\bibitem{Levin-TRG} M. Levin and C. P. Nave, Phys. Rev. Lett. \textbf{99}, 120601 (2007).

\bibitem{Jianghongchen-BVP} H. C. Jiang, Z. Y. Weng, and T. Xiang, Phys. Rev. Lett. \textbf{101}, 090603 (2008).

\bibitem{Guzhengcheng-TERG} Z. C. Gu, M. Levin, X. G. Wen, Phys. Rev. B \textbf{78}, 205116 (2008).

\bibitem{JOVVC-PEPSAlgor} J. Jordan, R. Orus, G. Vidal, F. Verstraete, and J. I. Cirac, Phys. Rev. Lett. \textbf{101}, 250602 (2008).

\bibitem{Xiezhiyuan-SRG} Z. Y. Xie, H. C. Jiang, Q. N. Chen, Z. Y. Weng, and T. Xiang, Phys. Rev. Lett. \textbf{103}, 160601 (2009).

\bibitem{Zhaohuihai-SRG} H. H. Zhao, Z. Y. Xie, Q. N. Chen, Z. C. Wei, J. W. Cai, and T. Xiang, Phys. Rev. B \textbf{81}, 174411 (2010).

\bibitem{Chen2011} Q. N. Chen, M. P. Qin, J. Chen, Z. C. Wei, H. H. Zhao, B. Normand, and T. Xiang, Phys. Rev. Lett. \textbf{107}, 165701 (2011)


\bibitem{Vers-SinLayer} I. Pizorn, L. Wang, F. Verstraete, Phys. Rev. A \textbf{83}, 052321 (2011).

\bibitem{Nishino-CTTRG} T. Nishino and K. Okunishi, J. Phys. Soc. Jpn. \textbf{67}, 3066 (1998).

\bibitem{Nishino-CTMRGKWA} K. Okunishi, T. Nishino, Prog. Theor. Phys, \textbf{103}, 541 (2000).

\bibitem{Nishino-TPVA} T. Nishino, K. Okunishi, Y. Hieida, N. Maeshima, Y. Akutsu, Nucl. Phys. B \textbf{575}, 504 (2000).

\bibitem{Nishino-CTMRGVertex} T. Nishino, Y. Hieida, K. Okunishi, N. Maeshima, Y. Akutsu, and A. Gendiar, Prog. Theo. Phys, \textbf{105}, 409 (2001).

\bibitem{Nishino-VDMA} N. Maeshima, Y. Hieida, Y. Akutsu, T. Nishino, and K. Okunishi, Phys. Rev. E \textbf{64}, 016705 (2001).

\bibitem{Nishino-TPVAPara} A. Gendiar, and T. Nishino, Phys. Rev. B \textbf{71}, 024404 (2005).

\bibitem{Latoree-BondScheme} A. Garcia-Saez, and J. I. Latorre, arXiv:1112.1412.

\bibitem{Wen-unpub} Z. C. Gu, M. Levin, and X. G. Wen, unpublished.

\bibitem{Roman-CTMReview} R. Orus, Phys. Rev. B \textbf{85}, 205117 (2012).

\bibitem{Latheauwer-HOSVD} L. de Lathauwer, B. de Moor, and J. Vandewalle, SIAM J. Matrix Anal. Appl, \textbf{21}, 1253 (2000).

\bibitem{Latheauwer-LowRank} L. de Lathauwer, B. de Moor, and J. Vandewalle, SIAM J. Matrix Anal. Appl, \textbf{21}, 1324 (2000).

\bibitem{Silva-LowRank} V. de. Silva, and L. H. Lim, SIAM. J. Matrix Anal. Appl. \textbf{30}, 1084 (2008).

\bibitem{HOSVD-AdvanApp} D. J. Luo, C. Ding, and H. Huang, arXiv:0902.4521; G. Bergqvist, E. G. Larsson, IEEE Signal Proc. Mag. \textbf{27}, 151 (2010).

\bibitem{White-DMRG} S. R. White, Phys. Rev. Lett. \textbf{69}, 2863 (1992).

\bibitem{note} In general, the bond density matrix may not be hermitian. In this case one should diagonalize this matrix using the canonical transformation, rather than the unitary transformation.

\bibitem{OnSager-RigSqIsing} L. Onsager, Phys. Rev. \textbf{65}, 117 (1944).

\bibitem{XiaomeiFeng-HeatCurveEc} X. M. Feng, and H. W. J. Blote, Phys. Rev. E \textbf{81}, 031103 (2010).

\bibitem{Fuji-SeriExpAlpha} H. Arisue, and T. Fujiwara, Phys. Rev. E \textbf{67}, 066109 (2003).

\bibitem{Martin-MCAlphaBeta} M. Hasenbusch, Int. J. of Mod. Phys. C \textbf{12}, 911 (2001).

\bibitem{Comi-SeriExpEc} P. Butera, and M. Comi, Phys. Rev. B \textbf{60}, 6749 (1999).

\bibitem{Fisher-SeriExpEc} A. J. Liu, and M. E. Fisher, Physica A \textbf{156}, 35 (1989).

\bibitem{Hunter-SeriExpEc} M. F. Sykes, D. L. Hunter, D. S. McKenzie, and B. R. Heap, J. Phys. A \textbf{5}, 667 (1972).

\bibitem{Pinn-MCEc} M. Hasenbusch, and A. Pinn, J. Phys. A \textbf{31}, 6157 (1998).

\bibitem{Jensen-MCEc} S. J. K. Jensen, and O. G. Mouritsen, J. Phys. A \textbf{15}, 2631 (1982).

\bibitem{Talapov-SponFormulaTc} A. L. Talapov, and H. W. J. Blote, J. Phys. A: Math. Gen. \textbf{29}, 5727 (1996).

\bibitem{Vicari-SeriExpBeta} M. Campostrini, A. Pelissetto, P. Rossi, and E. Vicari, Phys. Rev. E \textbf{65}, 066127 (2002).

\bibitem{Martin-MCTc} M. Hasenbusch, Phys. Rev. B \textbf{82}, 174433 (2010).

\bibitem{Dengyoujin-MCTc} Y. J. Deng, and H. W. J. Blote, Phys. Rev. E \textbf{68}, 036125 (2003).

\bibitem{Gupta-MCTc} R. Gupta, and P. Tamayo, Int. J. Mod. Phys. C \textbf{7} 305(1996).

\bibitem{Comi-SeriExpTc} P. Butera, and M. Comi, Phys. Rev. B \textbf{62}, 14837 (2000).

\bibitem{Chung-Tc} S. G. Chung, Phys. Lett. A \textbf{359}, 707 (2006).

\bibitem{Dengyoujin-MChc} H.W. J. Blote and Y. Deng, Phys. Rev. E \textit{66}, 066110 (2002).

\bibitem{Hamer-SeriExphc} H. X. He, C. J. Hamer, and J. Oitmaa, J. Phys. A \textbf{23}, 1775 (1990); J. Oitmaa, C. J. Hamer, and W. H. Zheng, J. Phys. A \textbf{24}, 2863 (1991).

\bibitem{Bursill96} R. J. Bursill, T. Xiang, G. A. Gehring, J. Phys.: Condensed Matter \textbf{8}, L583 (1996).

\bibitem{Wang97} X. Wang, T. Xiang, Phys. Rev. B \textbf{56}, 5061 (1997). Cond-mat/9705301.

\bibitem{Ran2012} S.J. Ran, W. Li, X. Xi, Z. Zhang, G. Su, arXiv:1205.5636, unpublished.

\bibitem{Roman-iPEPShc} R. Orus, G. Vidal, Phys. Rev. B \textbf{80} 094403 (2009).

\end{thebibliography}
\end{document}